# Measuring eye-tracking accuracy and its impact on usability in Apple Vision Pro

Zehao Huang[1], Gancheng Zhu[1], Xiaoting Duan[1], YongKai Li[1], Rong Wang[1], Shuai Zhang[2] &

Zhiguo Wang[1]

1 Center for Psychological Science, Zhejiang University, Hangzhou, China

2 Tianmushan Laboratory, Hangzhou, China

## Abstract

With built-in eye-tracking cameras, the recently released Apple Vision Pro (AVP) mixed reality (MR) headset features gaze-based interaction, eye image rendering on external screens, and iris recognition for device unlocking. One of the technological advancements of the AVP is its heavy reliance on gaze- and gesture-based interaction. However, limited information is available regarding the technological specifications of the eye-tracking capability of the AVP, and raw gaze data is inaccessible to developers. This study evaluates the eye-tracking accuracy of the AVP with two sets of tests spanning both MR and virtual reality (VR) applications. This study also examines how eye-tracking accuracy relates to user-reported usability. The results revealed an overall eye-tracking accuracy of 1.11° and 0.93° in two testing setups, within a field of view (FOV) of approximately 34° x 18°. The usability and learnability scores of the AVP, measured using the standard System Usability Scale (SUS), were 75.24 and 68.26, respectively. Importantly, no statistically reliable correlation was found between eye-tracking accuracy and usability scores. These results suggest that eye-tracking accuracy is critical for gaze-based interaction, but it is not the sole determinant of user experience in VR/AR.

# 1 Introduction

Eye-tracking is a technology used to measure the direction of gaze or the rotation of the eyeballs in humans and (or) animals, and various eye-tracking technologies have been used in research and industrial applications. The past two decades have seen an increasing number of video-based eye-tracking devices, thanks to the advancements in computer vision technology. Video-based eye-tracking is more user-friendly than other solutions, e.g., EOG and DPI. Yet, it can still provide high-precision eye movement data that meets the needs of both scientific and industrial applications. In recent years, eye-tracking has found its application in consumer-grade VR/AR headsets. Several popular VR devices, such as Meta Quest Pro, Pico 4 Pro, and Apple Vision Pro (AVP), have already incorporated eye-tracking functionality. Eye-tracking integration in VR/AR devices helps improve graphic computation, e.g., foveated rendering (Kim et al., 2019; Meng et al., 2020; Patney et al., 2016) and varifocal display (D. Dunn, 2019). Eye-tracking also helps to enhance the interactive experience in VR/AR (Gardony et al., 2020; Nam & Choi, 2023).

## 1.1 Eye-tracking offers a new interaction modality

Eye-tracking has been widely used in medical and healthcare fields (Chatelain et al., 2020; Larsson et al., 2015). For instance, eye trackers allow patients with conditions like Amyotrophic Lateral Sclerosis (ALS) to communicate through gazing on-screen interactive elements (J. Wang et al., 2023). However, eye-tracking as an interaction modality is still scarce in commercial VR/AR headsets. The input methods used in VR devices are still limited, with many VR/AR devices relying on controllers (e.g., Meta Quest 2, 3; Pico Neo 2) or head-turning (e.g., HoloLens 1) to interact with elements. Fernandes and his colleagues (2023) compared user acceptance of three interaction methods and found that eye-tracking and controllers received similar ratings on interaction efficiency and usability, both higher than head-turning. However, in this study, the miss rate of eye-tracking was significantly higher than that of the other two interaction methods.

There is a wealth of research works and prototypes of eye-tracking for interaction in VR/AR devices (Pai et al., 2019; Sipatchin et al., 2021; Wei et al., 2023). However, no product has abandoned controllers entirely, relying solely on eye-tracking and gesture control to operate the device. Menu navigation, object interaction, and other operations are still performed using controllers, with eye-tracking as supplementary input (Adhanom et al., 2023). In Pico 4 Pro, eye-tracking is used to measure pupil position only. In PSVR 2, eye-tracking can be used for foveated rendering or as an auxiliary input device for gaming. The limited use of eye-tracking in VR/AR devices may be partially due to the insufficient performance (accuracy, precision, data loss, etc.) of the eye trackers available for VR/AR devices. The accuracy of the eye-tracking devices integrated in VR headsets is generally lower than that of head-mounted eye trackers (HMET)[1]. Furthermore, using eye-tracking for interaction requires significant modifications to the user interface (Feit et al., 2017; Pfeuffer et al., 2017) and a steep learning curve for users (Akiyoshi & Takeno, 2013; Valtakari et al., 2021).

**1.2 Eye-tracking for foveated rendering**

Another primary use of eye-tracking in VR/AR devices is foveated rendering. To provide an immersive visual experience, the resolution of head-mounted displays (HMDs) continues to increase in state-of-the-art VR/AR devices, e.g., from 2.3 million pixels per eye in Oculus Rift (2019) to 14 million pixels per eye in Varjo XR-4 (2024). VR/AR devices equipped with eye trackers usually support foveated rendering (Mohanto et al., 2022), i.e., images rendered on the HMDs have high resolution only in the central visual field of the user, but the peripheral visual field is rendered in lower resolution (see Figure 1, for an illustration). Foveated rendering effectively reduces the computational load on VR/AR devices (Nyamtiga et al., 2024), it also aligns with the characteristics of the human visual system (L. Wang et al., 2023).

---

[1] The eye-tracking devices used in VR/AR typically have similar sensor layouts as head-mounted eye trackers (HMET). However, with a small eye relief, eye-tracking cameras are not placed in optimal tracking positions in VR/AR devices. As documented in the literature (Ehinger et al., 2019; Pastel et al., 2021; Sipatchin et al., 2021), the eye-tracking accuracy of VR/AR eye trackers is typically 2-4°, much lower than HMET.

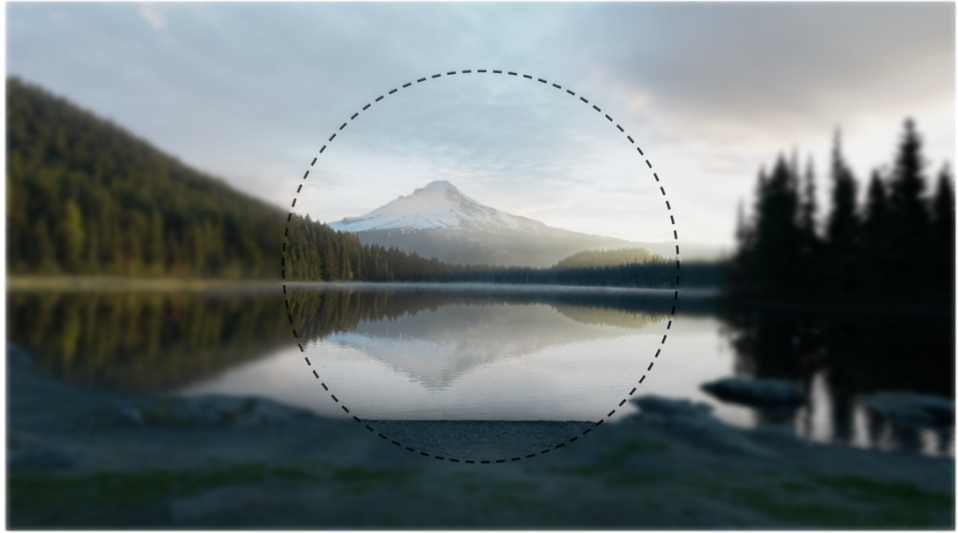

*Figure 1 A screenshot from AVP. Image elements inside the dashed circle are rendered clearer than those outside the circle.*

**1.3 Eye-tracking in Apple Vision Pro**

Apple Inc. introduced gaze interaction to its VR/AR (or mix-reality device, MR) device, Apple Vision Pro (AVP) in late 2023, setting up a new stage for eye-tracking in VR/AR. In AVP, gaze data provided by the eye tracker is linked to the current area of interest of the user, such as a menu item or an icon, like the function of a mouse cursor. Users select elements that may require action through gaze and then perform further actions on the selected icon, such as confirmation or zooming. Furthermore, gaze data from the eye tracker is utilized for foveated rendering, to reduce visual fatigue or vertigo (Koulieris et al., 2019; Mohanto et al., 2022; Zhou et al., 2021). Unlike other VR/AR devices, AVP deprecated controllers in favor of a new interactive technology that relies on eye-tracking and hand-tracking, showing that Apple has strong confidence in its VR/AR eye-tracking device.

**1.4 Aims of the present study**

The primary objective of this study is to evaluate the eye-tracking accuracy in AVP. There is no publicly available data on eye-tracking accuracy in AVP. However, based on previous studies examining the eye-tracking data quality of research-grade eye trackers (e.g., Aziz & Komogortsev, 2022; Brand et al., 2021; Cuve et al., 2022; Ehinger et al., 2019; Hessels et al., 2015; Holmqvist & Blignaut, 2020; Hooge et al., 2022; Morgante et al., 2012; Pastel et al., 2023; Schuetz & Fiehler, 2022), the tracking accuracy of AVP is unlikely to surpass the gold standard EyeLink eye trackers, which is about 0.3° in optimal operation conditions. Nevertheless, measuring the eye-tracking accuracy of AVP could provide developers with a good reference for VR/AR eye-tracking applications.

AVP developers cannot directly access the eye-tracking data, for privacy considerations. The present study estimated the gaze position with two different testing pipelines. The first testing pipeline leverages the accessibility features of the AVP, enabling users to display a gaze pointer that can be recorded during a screen capture. This recording can later be analyzed with computer vision algorithms to estimate gaze position. The second testing pipeline leverages a Unity functionality, Unity PolySpatial Input, with which an input event is registered when the user is looking at an object and performing a pinch gesture with the index finger and thumb. For convenience, we will refer to these two testing pipelines as "accessibility" and "unity" pipelines, respectively.

Furthermore, although eye-tracking and hand gestures are not new interaction technologies in VR/AR (Pfeuffer et al., 2017), AVP is the first to abandon controllers and rely solely on eye-tracking and hand gestures. Whether this novel interaction method poses a great learning burden to users is yet documented. In addition to testing eye-tracking accuracy, we also evaluated the user-reported usability of AVP to gain a more comprehensive picture of eye-tracking in VR/AR interaction.

## 2 Methods

The research protocol reported in this paper was approved by a local ethics committee at the Center for Psychological Sciences at Zhejiang University, and all volunteers provided written informed consent.

As in previous studies that examined the accuracy of eye-tracking devices or algorithms (Hessels et al., 2015; Holmqvist et al., 2012; Huang et al., 2024; Macinnes et al., 2018), a fixation task in which the volunteer looked at visual targets at pre-specified locations was tested. As noted above, two different testing pipelines were used in the present study, leveraging the accessibility features of AVP (the accessibility pipeline) and Unity PolySpatial Input (the unity pipeline). The accessibility pipeline allows us to estimate gaze position continuously at a speed of 30 frames per second (FPS), however, the visibility of the gaze pointer (Figure 2B) may encourage the volunteers to adjust their gaze involuntarily. The unity pipeline does not have this issue, however, only one gaze sample is recorded when a pinch gesture occurs. The present study adopted both testing pipelines to gain a more comprehensive estimate of the eye-tracking accuracy of the AVP.

In addition to eye-tracking accuracy, the present study also assessed the perceived usability of AVP using the System Usability Scale (Brooke, 1996).

### 2.1 Volunteers

Twenty-seven volunteers (11 females/ 16 males, age: M = 24.59 years, SD = 2.72) completed the fixation task in both testing pipelines and filled out the SUS questionnaire. One volunteer was excluded from the analysis due to a device issue.

## 2.2 Equipment and tools

### 2.2.1 The AVP headset

The AVP used in the test featured micro-OLED displays with 23 million pixels and refresh rates of 90Hz, 96Hz, and 100Hz. It had four eye-tracking cameras for foveated rendering and external cameras for spatial awareness and hand tracking. Equipped with a dual-chip processor, the AVP provides efficient computational power. Additionally, the device operates on the newly developed visionOS (system version 1.0.3) and weighs approximately 600-650 grams.

### 2.2.2 Usability rating scale

The System Usability Scale (SUS) is widely used for measuring system usability (Lewis, 2018). With only 10 items, the SUS is convenient to administer but still has high reliability and sensitivity, even on small samples (Bangor et al., 2008). The Chinese version of SUS has the same reliability and validity as the original version (Y. Wang et al., 2020). This tool was used to evaluate the overall user experience of AVP, which replaces the controllers typically seen in VR devices with gaze and hand gesture interaction.

## 2.3 Testing setup

The eye-tracking accuracy of the AVP was assessed in a fixation task, in which the volunteers looked at visual targets at pre-specified positions. The fixation task was tested in both accessibility and unity pipelines.

### 2.3.1 The accessibility pipeline

When tested in the accessibility pipeline, the volunteer rested their chin on a chinrest. A physical screen (24-inch, 1080p, 165 Hz), placed 80cm from the chinrest, was used to present the fixation targets (see Figure 3 A). Following previous work that evaluated the eye-tracking accuracy of head-mounted eye trackers (Ehinger et al., 2019; Huang et al., 2024), 16 ArUco markers were constantly presented at the edge of the screen area (see Figure 2A), and these markers were later used to extract the screen area using a Gaze Mapping Algorithm (Macinnes et al., 2018). Nine fixation targets would appear on the screen along the ArUco markers, spanning a field of view (FOV) of $34° \times 18°$ (as shown in Figure 2A). The presentation of the ArUco markers and fixation targets was controlled by custom scripts written in PsychoPy2 (Peirce et al., 2019). The fixation target would randomly appear at one of 9 pre-specified screen positions. The volunteers were required to fixate on the target for at least 2 seconds, then press the space bar to move the target to the next position. The fixation target appeared at each of the 9 fixation target positions 4 times.

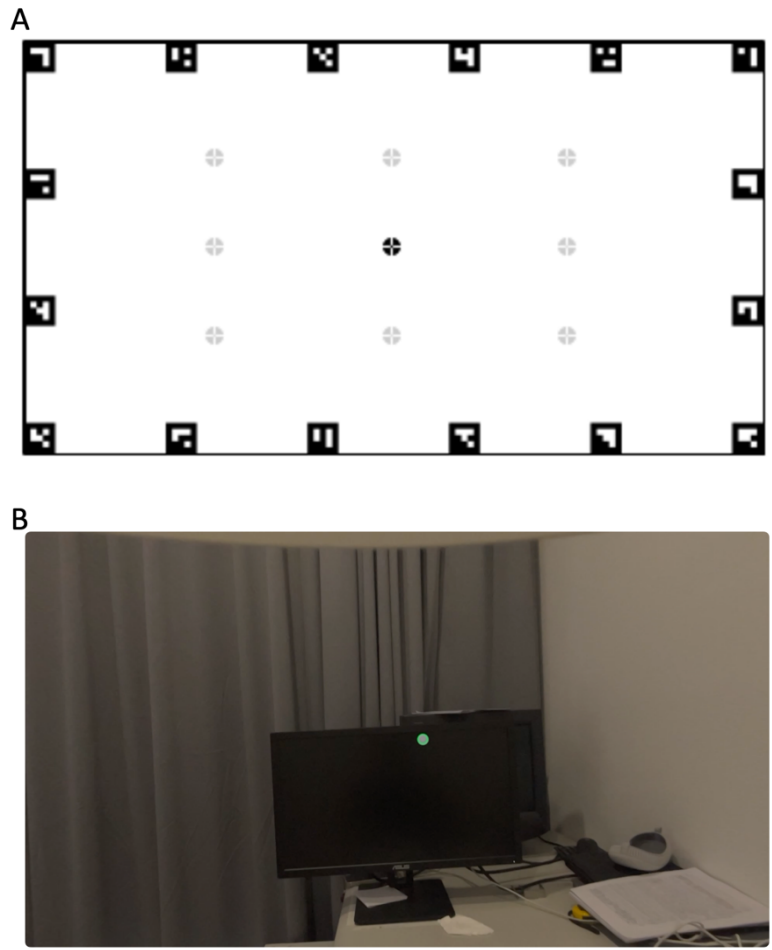

*Figure 2 (A) Schematic diagram of the external screen experimental program. (B) AVP screenshot, with the green circle in the center representing the Pointer Controller.*

The accessibility pipeline was performed in the see-through mode so the volunteers could see and interact with the external world. During the test, the volunteer first changed the Pointer Control setting of AVP to display a real-time gaze pointer (green disk) on the AVP screen (see Figure 2B). Then, the volunteer turned on screen capture until the fixation task was completed. This screen capture is much like the scene camera data of head-mounted eye trackers, in which a gaze cursor overlays the captured scene. The position of the pointer can then be extracted using computer vision algorithms.

### 2.3.2 The Unity pipeline

The Unity pipeline used Unity (version: 2022.3.27f1c1) PolySpatial visionOS API (Version 1.1) to retrieve the gaze position while the AVP was set to operate in the immersion mode. Like the accessibility pipeline, a chinrest was used to minimize head movements (see Figure 3B). In this test, the fixation targets appeared at 9 positions on a virtual panel, covering a 34° × 18° FOV. The volunteer was instructed to pinch the index finger and thumb while looking at the visual target when it appeared on the virtual panel. Then, the target was displaced to a different position and the volunteer followed the displacement and pinched their fingers. After the visual target had appeared at each of the target positions 4 times, the testing program would automatically exit.

### 2.3.3 Task procedure

All volunteers were tested following the protocol presented below (Figure 3C).

(1) A 10-minute introduction and training session to familiarize the volunteers with the functionality and features of the AVP.

(2) Eyes and hands setup in the AVP (eye-tracking and hand-tracking calibration).

(3) The volunteer placed his/her chin on the chinrest and completed the fixation task in the accessibility and Unity testing pipelines. The order of these two testing pipelines was counterbalanced across the volunteers.

(4) The volunteer was given another 10 minutes to freely explore the various features and functions of the AVP.

(5) The volunteers removed the AVP to fill out the SUS questionnaire.

(6) The volunteers put on the AVP again and completed the fixation task in the Unity pipeline for a second time, without recalibrating eye-tracking and hand-tracking.

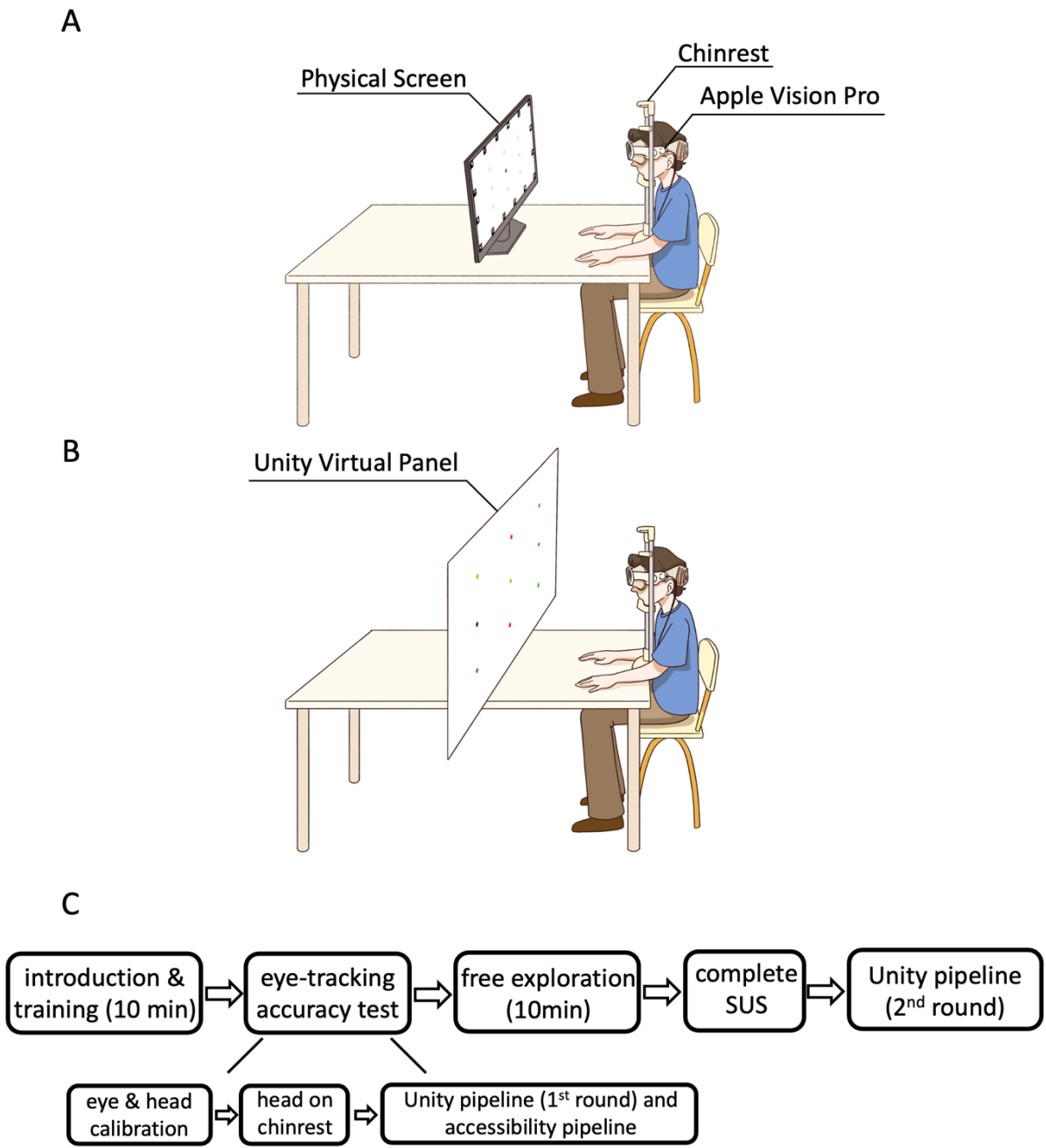

*Figure 3 (A) Testing setup in the accessibility pipeline. (B) Testing setup in the Unity pipeline. (C) The task procedure of a single testing session.*

## 2.4 Data analysis

### 2.4.1 Gaze data in the accessibility pipeline

The output of the accessibility pipeline is a screen capture with the gaze cursor (pointer) overlay on the scene (see Figure 2B). We used a circle feature matching algorithm in OpenCV (version 4.9.0) to estimate the center of the pointer (i.e., the gaze position) in each frame. This method helped us to extract gaze coordinates at 30 FPS. We then used a gaze mapping algorithm (MacInnes et al., 2018) to convert the scene-based gaze coordinates to screen-based coordinates.

To align the time series gaze data with the visual target, we used Adobe Premiere (version 24.5.0) to manually segment the frames of the AVP screen capture based on the position of the fixation target. For convenience, we will refer to the time window in which the visual target appeared at a given position as an "epoch". For each epoch, the gaze coordinates from three continuous frames (~ 100 ms) were selected to estimate the eye-tracking accuracy (Huang et al., 2024; for an in-depth discussion on sample selection and eye-tracking data accuracy, see Holmqvist et al., 2012). With an *n*-sample recording *T*, the samples for estimating tracking accuracy are selected using a sliding window with a stride of 1 and a window size of $k$ ($k = 3$, in the present study). The sliding window will generate a set of ($n - k + 1$) segments of consecutive samples, $S = \{s_1, s_2, s_3, \ldots, s_{n-k+1}\}$. Gaze samples are then selected by finding a segment of minimum position variance (STD) and deviation from the ground truth (DEV), the metric $Selected\ Samples = \underset{i \in \{1,2,3\ldots,n-k+1\}}{\mathrm{argmin}} (DEV(s_i)^2 \times STD(s_i)^2)$, where $\forall s_i: s_i \in T$. Then the eye-tracking accuracy was quantified with the mean absolute error (MAE), i.e., the average gaze deviation from the visual target.

### 2.4.2 Gaze data in the Unity pipeline

In the Unity pipeline, a gaze position was recorded when the volunteer pinched his/her index finger and thumb. The eye-tracking accuracy was quantified with the gaze deviation from the visual target in 3D space.

### 2.4.3 SUS ratings

For the SUS, we examine the overall rating and the ratings on the learnability and usability subscales. Learnability is assessed using items 4 and 10 of the SUS, while usability is evaluated using the remaining items.

## 3 Results

### 3.1 Eye-tracking accuracy

In the fixation task used to assess the eye-tracking accuracy of AVP, the volunteer looked at a fixation target at 9 pre-specified positions on a physical screen in the accessibility pipeline, or on a virtual panel in the Unity pipeline (see Figures 3A & 3B). The primary variable of interest is eye-tracking accuracy, i.e., the distance between the gaze position reported by the AVP and the actual fixation target position. For easy comparison to previous studies on eye-tracking data quality, we reported the eye-tracking accuracy in degrees of visual angle (°).

The testing results reported here are based on data from 26 volunteers.

### 3.1.1 Eye-tracking accuracy in the accessibility pipeline

In the accessibility pipeline, the average eye-tracking accuracy was 1.11° (SD = 0.77, range: 0.13°-7.58°). The eye-tracking accuracy distributions at the 9 target positions are presented in Figure 4B (see also Table A1 in the Appendix).

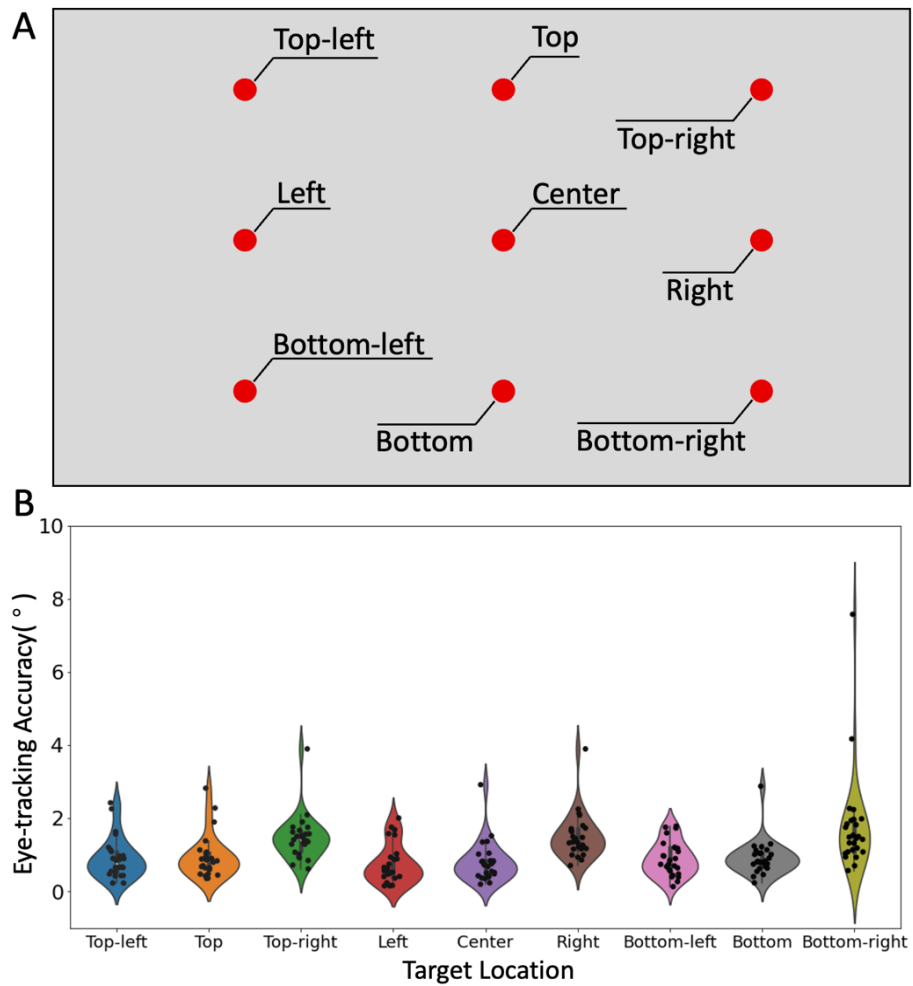

*Figure 4 (A) The layout of the nine visual targets. (B) The eye-tracking accuracy at each target position, as revealed by tests in the accessibility pipeline.*

### 3.1.2 Eye-tracking accuracy in the Unity pipeline

In the Unity pipeline, the volunteers removed the AVP from their heads following the first round of tests. Then, they put on the headset without recalibrating the eye-tracker and completed the tests a second time. The eye-tracking accuracy distributions at each target location are presented in Figure 5 (see also Table A2 in the Appendix). The average eye-tracking accuracy of the two rounds of tests was 0.7° (SD = 1.00, range: 0.10° – 8.55°) and 1.15° (SD = 1.80, range: 0.09° - 10.61°), respectively.

An ANOVA on the eye-tracking accuracy, with variables target position and testing rounds as within-subjects factors, revealed a marginally significant main effect for target position [$F(8,43) = 2.72$, $p = 0.06$] and testing rounds [$F(1,50) = 2.831$, $p = 0.099$]. The two-way interaction between target position and testing rounds was not statistically significant [$F(8,43) = 1.22$, $p = 0.31$]. The marginal effect of target position appeared because all camera-based tracking solutions tend to perform better for the central FOV. While the AVP can maintain good eye-tracking accuracy without recalibration, the eye-tracking accuracy did degrade by 0.45° in the second round of tests.

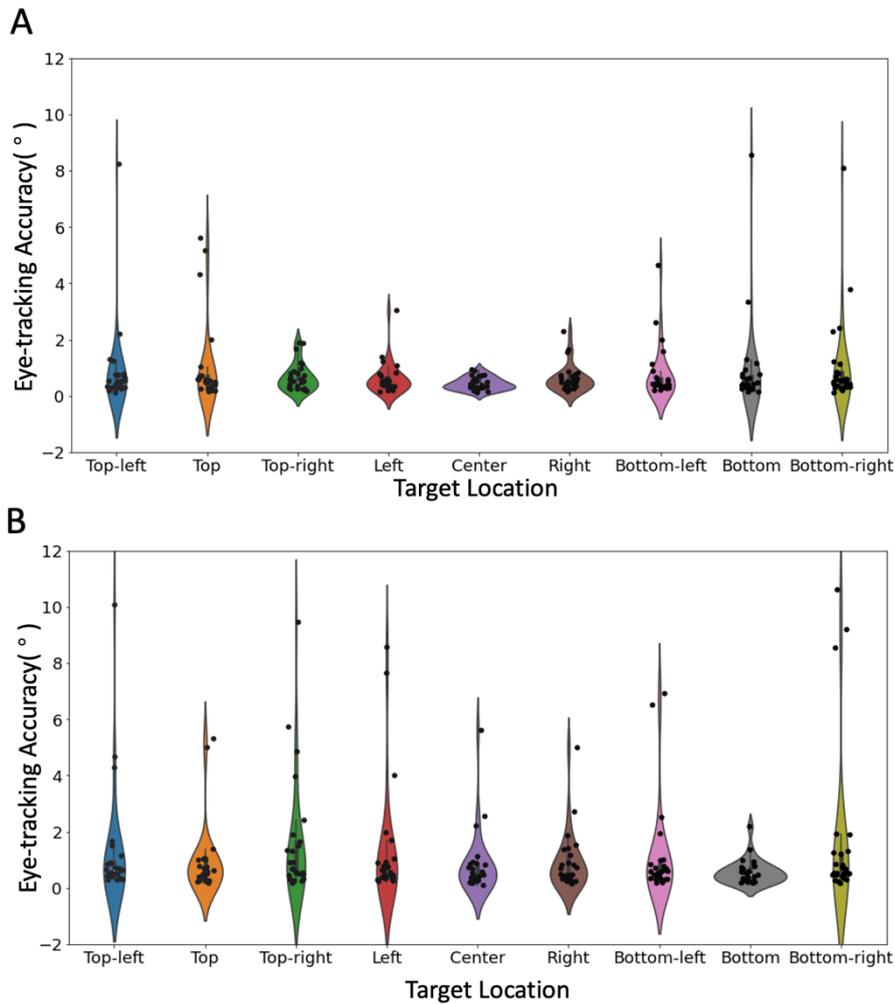

*Figure 5 Eye-tracking accuracy in the first (A) and second (B) Unity pipeline test.*

### 3.2 SUS ratings

The SUS ratings were converted to standard scores and then aggregated separately on the usability and learnability subscales. Compared to the industrial norm of SUS (Lewis & Sauro, 2018), the average usability score for AVP is 74.56 (SD =10.77). The learnability of the AVP score is 69.50 (SD = 22.66). The SUS overall score is 73.55 (SD = 10.61), falling within grade B-, higher than 65-69% of products. The SUS scores for each volunteer are presented in the Appendix (Table A3).

To examine how eye-tracking accuracy impacts usability, we calculated the Pearson correlation between the gaze error of the first-round Unity pipeline test and the SUS ratings. The results revealed no statistically reliable correlation (see Figure 6) between the overall gaze error and the two SUS subscales, usable ($r = 0.084$, $p = 0.684$) and learnable ($r = 0.061$, $p = 0.766$), as well as the overall SUS score ($r = 0.097$, $p = 0.638$). These results suggest that, while the eye-tracking accuracy of the AVP varied among users, eye-tracking accuracy did not significantly impact user experience. [2]

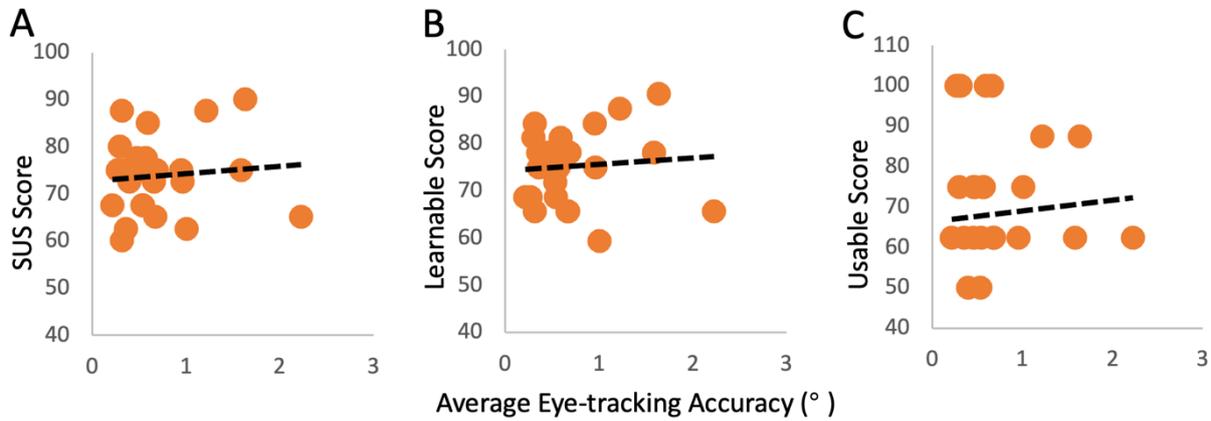

*Figure 6 Scatter plots showing the correlation between eye-tracking accuracy and SUS score (A), and learnable (B) and usable (C) scores.*

## 4 Discussion

### 4.1 The eye-tracking accuracy of the AVP

The primary objective of this study was to evaluate the eye-tracking accuracy of the AVP. We used a standard fixation task, in which the volunteers looked at 9 fixation targets on a physical screen in the see-through mode (accessibility pipeline) or a virtual panel in the immersive mode (Unity pipeline). Apple Inc. does not allow developers to access the raw eye-tracking data of the AVP. Thus, we estimated the gaze position with the center of the pointer in the accessibility pipeline and the Unity PolySpatial Input in the Unity pipeline. The testing results indicate that the overall eye-tracking accuracy of AVP is 1.10° in the accessibility pipeline test and 0.93° in the Unity pipeline. Furthermore, there is no correlation between eye-tracking accuracy and user-reported usability of AVP.

In research applications, eye-tracking data quality is evaluated with multiple metrics, e.g., accuracy, precision, latency, and data loss (M. J. Dunn et al., 2023; Holmqvist et al., 2012). The same set of data quality metrics also applies to VR/AR eye-tracking. For example, previous studies have shown that, with foveated rendering, the latency of the gaze signal should not exceed 70 ms (Albert et al., 2017). For eye-tracking accuracy, Sipatchin and her colleagues (2021) tested an HTC Vive Pro Eye and revealed that the eye-tracking accuracy of the built-in Tobii eye-tracking device varied between 0.94° to 10.77° (mean =

---

[2] We note that the lack of correlation between the SUS scores and tracking accuracy may be due to the low variability of eye-tracking accuracy in the tested population (see Figures 4 & 5) in the present study. Eye-tracking accuracy may not significantly impact user experience as long as it remains within an acceptable range. The correlation reported here should be interpreted with caution.

4.16°) over an FOV of 48° × 48°. Schuetz & Fiehler (2022) performed a similar test on the same device and reported an overall eye-tracking accuracy between 0.09° and 4.99° (mean = 1.08°). Both studies have shown that the central FOV enjoys the highest eye-tracking accuracy, whereas eye-tracking accuracy decreases moving towards more peripheral regions. From Figure 4B and Figure 5, we can also clearly find that the eye-tracking accuracy of the AVP is highest in the central FOV, which is on par with or better than other VR devices (Schuetz & Fiehler, 2022b; Sipatchin et al., 2021; for a review, see Clay et al., 2019).

Head-mounted eye-tracking glasses predate VR eye-tracking. However, HMET and VR eye-tracking solutions are similar in form, structure, and tracking algorithms. Apple Inc. acquired SensoMotoric Instruments (SMI) in 2017; the eye-tracking solution implemented in the AVP may be based on SMI eye-tracking technology. Previous studies have shown that the gaze error of the SMI-ETG2 eye-tracking glasses ranges between 0.55° (as reported by Pastel et al., 2021) to 1.21° (as reported by Huang et al. 2024). The placement of the infrared cameras in the AVP may differ from SMI-ETG2 due to the smaller eye relief in the AVP. However, the present study showed that the eye-tracking accuracy of the AVP is about 1.1°, which is comparable to SMI-ETG2 and other top-tier head-mounted eye-tracking glasses (Ehinger et al., 2019; Hooge et al., 2022; Huang et al., 2024; Niehorster, Santini, et al., 2020; Pastel et al., 2021).

**4.2 Eye-tracking accuracy and gaze-based interaction in VR/AR**

Previous studies have extensively explored the possible application of eye-tracking in interaction, however, few studies have examined the eye-tracking accuracy needed for different types of gaze-based interactions (Feit et al., 2017; Majaranta & Bulling, 2014; Valtakari et al., 2021). Previous studies have shown that large icons and menus facilitate quicker and more accurate gaze-based selection (Kammerer & Beinhauer, 2010; Niu et al., 2021; Yi-yan et al., 2024); however, they may also impose serious space restrictions on UI design. While small icons save space, they make interaction challenging and diminish the user experience. Feit and her colleagues (2017) conducted a benchmark test of multiple eye-tracking devices for daily input, and they found a significant decrease in eye-tracking accuracy and precision in screen corners. Therefore, they suggested that UI design should adaptively adjust the size of UI elements based on their positions. For example, icons at the bottom of the screen should be as large as possible.

Gaze-based interaction is critical to the AVP, which deserted controllers in favor of gaze and gesture interaction. The Apple Design Guidelines recommend that the gaze target be at least 60 points. A circular icon may have a diameter of 44 points, but the associated interactive area is larger, forming a square with a side length of 60 points. Apple also recommended that the center-to-center distance between two icons be at least 60 points (*Designing for Spatial Input*, 2023). An icon of 60 points corresponds to a visual angle of 0.7°. This design guideline agrees with the present testing results in the Unity pipeline. However, the average eye-tracking accuracy of the AVP dropped to 1.10° when the volunteer removed and then put the AVP back on without recalibration. Some volunteers verbally reported that selecting menu items was not as effortless as before, requiring more effort. These observations suggest that when designing eye-tracking interaction elements for VR/AR devices, accuracy in eye-tracking is critical. The present testing results are a good reference for other VR/AR manufacturers to evaluate their eye-tracking solutions.

**4.3 Eye-tracking accuracy and usability**

Fernandes et al. (2023) compared three target selection methods and found that controllers were the most precise, but eye-tracking input was more natural and easier to use, and both are better than head-turning. Luro & Sundstedt (2019) also found that gaze-based selection significantly reduces cognitive load. However, previous studies have shown that gaze-based input selection may decrease interaction accuracy (Qian & Teather, 2017). Given that the gaze error of most eye-tracking glasses is greater than 1.0° (Huang et al., 2024; Lamb et al., 2022; Niehorster, Hessels, et al., 2020; Sipatchin et al., 2021), low eye-tracking accuracy may be a hurdle to usability in gaze-based interactions.

The volunteers in the present study could accurately select and manipulate menu items while they performed the tests or freely explored the AVP most of the time. The testing results also showed that the eye-tracking accuracy of AVP does not appear to impact usability. These findings indicate that the usability of gaze-based interaction is not determined by eye-tracking accuracy but rather heavily influenced by interaction design. For example, previous work has shown that well-designed feedback allows users to quickly and easily identify their currently selected UI elements during gaze-based interaction (Kumar et al., 2020; Menges et al., 2019; Zhao et al., 2023). Many volunteers reported difficulty triggering the "close" button located at the bottom FOV of the AVP in the present study, likely due to the lower tracking accuracy at the bottom compared to the center FOV. However, the feedback design of AVP allowed the volunteers to learn that they did not select a target, and then adjust their head orientation or gaze to select the button. These adjustments in gaze direction and head orientation are mostly spontaneous behaviors and do not cause much confusion among the users.

**4.4 Limitations of the present work**

Before we conclude this paper, it is important to note that the present tests did not capture continuous gaze data directly from the AVP eye-tracking API. Instead, two methods (testing pipelines) were used to estimate the gaze position in the AVP. The first method leverages an accessibility feature of the AVP, which allows us to capture the AVP screen with a real-time gaze pointer at 30 FPS. The second method extracted the gaze position using a new SpatialPointerDevice in Unity, which returns the gaze position when the user pinches the index finger and thumb.

It remains unclear how well the mixed reality environment of AVP accurately captures the real physical environment (Chen et al., 2014; Hong, 2022). The viewing distance used to calculate visual angles in the accessibility pipeline may inflate the estimated eye-tracking accuracy of AVP. Furthermore, due to the API access restrictions imposed by Apple Inc., we could not evaluate other eye-tracking data quality measures that may heavily impact the user experience of gaze-based interaction, most notably precision, and latency.

## 5 Conclusion

The eye-tracking module of the recently released Apple Vision Pro mixed reality device is exclusively used for gaze-based interaction and graphics rendering, etc. To address privacy concerns, AVP does not allow developers to access its raw eye-tracking data. It remains unclear whether Apple will eventually open its eye-tracking interface to developers for applications in education, clinical research, and other fields. The current study evaluated the tracking accuracy of the eye-tracking module in the AVP. The test results revealed an overall eye-tracking accuracy of 0.93°-1.10° over an FOV of approximately 34° × 18°. This level of tracking accuracy is superior to research-grade eye-tracking glasses, such as the SMI-ETG2, Tobii Pro Glasses 2, and AdHawk MindLink (see Huang et al., for a recent test). Interestingly, we found that the overall usability ratings do not vary much with eye-tracking accuracy, showing that an overall tracking accuracy of approximately 1° does not significantly impact the user experience. We conclude that future research on VR/AR eye-tracking should prioritize interaction design over eye-tracking accuracy.

## 6 Open practices statement

*Data, experimental scripts, and analysis tools are available for research purposes and can be accessed from https://pan.baidu.com/s/1QxvT9iryhL-8jMkGHyCGLQ?pwd=xwy3. Passcode: xwy3.*

# 8 Appendix

Table A1 results of each volunteer in the accessibility pipeline test. Note: Units are visual angle, values in parentheses representing standard deviation.

| No | Bottom-left | Bottom | Bottom-right | Left | Center | Right | Top-left | Top | Top-right | Mean (SD) |
|---|---|---|---|---|---|---|---|---|---|---|
| 01 | 0.42 | 0.60 | 1.63 | 0.66 | 0.83 | 2.09 | 0.85 | 1.30 | 1.97 | 1.15 (0.59) |
| 02 | 0.67 | 1.14 | 1.47 | 0.15 | 0.82 | 1.64 | 0.48 | 0.72 | 2.24 | 1.04 (0.61) |
| 03 | 0.98 | 0.87 | 1.76 | 2.01 | 1.05 | 1.80 | 1.79 | 1.22 | 1.54 | 1.45 (0.40) |
| 04 | 0.66 | 0.36 | 1.29 | 0.47 | 0.35 | 1.00 | 0.68 | 0.64 | 0.94 | 0.71 (0.30) |
| 05 | 2.43 | 2.82 | 3.90 | 1.55 | 2.92 | 3.90 | 1.73 | 2.88 | 4.17 | 2.92 (0.88) |
| 06 | 0.96 | 0.64 | 1.02 | 0.41 | 0.43 | 1.73 | 1.32 | 0.80 | 0.70 | 0.89 (0.40) |
| 07 | 1.09 | 1.38 | 1.58 | 0.53 | 0.50 | 1.25 | 0.44 | 0.55 | 1.04 | 0.93 (0.41) |
| 08 | 0.67 | 0.87 | 1.57 | 0.24 | 0.68 | 1.16 | 0.39 | 0.80 | 1.32 | 0.86 (0.41) |
| 09 | 0.55 | 1.05 | 1.90 | 0.17 | 0.84 | 2.25 | 0.70 | 1.01 | 2.27 | 1.19 (0.72) |
| 10 | 0.90 | 0.71 | 1.52 | 0.88 | 1.02 | 1.00 | 0.60 | 0.24 | 1.19 | 0.89 (0.34) |
| 11 | 0.23 | 0.47 | 0.92 | 0.41 | 0.37 | 1.27 | 0.75 | 1.01 | 1.82 | 0.81 (0.48) |
| 12 | 1.14 | 0.80 | 1.51 | 0.63 | 0.24 | 1.48 | 1.09 | 0.83 | 1.49 | 1.02 (0.41) |
| 13 | 0.23 | 0.68 | 1.41 | 0.75 | 0.55 | 1.44 | 0.97 | 1.04 | 1.99 | 1.01 (0.51) |
| 14 | 0.43 | 0.95 | 1.29 | 0.15 | 0.84 | 1.22 | 0.27 | 0.81 | 1.50 | 0.83 (0.44) |
| 15 | 0.65 | 0.89 | 0.95 | 0.52 | 0.72 | 0.94 | 0.13 | 0.46 | 0.93 | 0.69 (0.26) |
| 16 | 0.66 | 0.35 | 0.85 | 1.03 | 0.39 | 1.34 | 1.19 | 1.14 | 1.84 | 0.98 (0.45) |
| 17 | 1.57 | 1.90 | 2.10 | 1.76 | 1.52 | 2.14 | 0.90 | 1.12 | 1.75 | 1.64 (0.39) |
| 18 | 0.48 | 0.45 | 0.62 | 0.94 | 0.48 | 1.67 | 1.76 | 1.05 | 7.58 | 1.67 (2.14) |
| 19 | 2.26 | 2.28 | 1.67 | 1.69 | 1.37 | 1.33 | 1.21 | 0.91 | 0.57 | 1.48 (0.54) |
| 20 | 1.20 | 1.05 | 1.37 | 0.89 | 0.76 | 1.35 | 0.64 | 0.41 | 1.33 | 1.00 (0.33) |
| 21 | 0.86 | 0.90 | 1.31 | 0.36 | 0.52 | 1.17 | 0.74 | 0.63 | 1.14 | 0.85 (0.30) |
| 22 | 0.43 | 0.48 | 1.41 | 0.56 | 0.76 | 1.50 | 1.18 | 0.69 | 1.46 | 0.94 (0.42) |
| 23 | 0.88 | 0.42 | 1.07 | 0.59 | 0.20 | 0.86 | 0.98 | 0.75 | 1.08 | 0.76 (0.29) |
| 24 | 1.64 | 0.84 | 0.72 | 1.58 | 0.82 | 0.71 | 1.60 | 0.76 | 1.08 | 1.08 (0.38) |
| 25 | 0.53 | 1.08 | 1.76 | 0.21 | 1.35 | 1.71 | 0.40 | 1.24 | 1.94 | 1.14 (0.60) |
| 26 | 0.91 | 0.64 | 1.36 | 0.52 | 0.42 | 1.17 | 0.56 | 0.83 | 1.43 | 0.87 (0.35) |
| Mean (SD) | 0.90 (0.54) | 0.95 (0.58) | 1.46 (0.6) | 0.76 (0.53) | 0.8 (0.54) | 1.50 (0.61) | 0.90 (0.46) | 0.92 (0.47) | 1.78 (1.35) | |

Table A2 Results of each volunteer in the Unity pipeline.

| No | Bottom-left | Bottom | Bottom-right | Left | Center | Right | Top-left | Top | Top-right | Mean (SD) |
|---|---|---|---|---|---|---|---|---|---|---|
| *1st round* | | | | | | | | | | |
| *Unity pipeline* | | | | | | | | | | |
| 01 | 0.52 | 0.47 | 0.47 | 0.36 | 0.29 | 0.25 | 0.32 | 0.30 | 0.24 | 0.36 (0.10) |
| 02 | 0.28 | 1.02 | 0.79 | 0.52 | 0.73 | 0.72 | 0.35 | 0.33 | 0.59 | 0.59 (0.23) |
| 03 | 0.43 | 0.70 | 0.71 | 0.33 | 0.52 | 0.37 | 0.28 | 0.30 | 0.48 | 0.46 (0.15) |
| 04 | 2.19 | 1.99 | 1.66 | 1.37 | 0.88 | 1.54 | 1.98 | 0.31 | 8.09 | 2.22 (2.15) |
| 05 | 0.53 | 0.39 | 0.25 | 0.45 | 0.48 | 0.38 | 0.38 | 0.59 | 0.67 | 0.46 (0.12) |
| 06 | 0.24 | 0.19 | 0.17 | 0.20 | 0.14 | 0.14 | 0.19 | 3.33 | 0.19 | 0.53 (0.99) |
| 07 | 0.64 | 0.53 | 0.79 | 0.58 | 0.69 | 0.75 | 0.64 | 0.75 | 0.79 | 0.68 (0.09) |
| 08 | 0.75 | 5.60 | 1.86 | 0.55 | 0.74 | 1.63 | 1.57 | 1.15 | 0.41 | 1.59 (1.50) |
| 09 | 0.27 | 0.18 | 0.33 | 0.19 | 0.31 | 0.43 | 0.39 | 0.23 | 0.35 | 0.30 (0.08) |
| 10 | 0.39 | 0.46 | 0.24 | 0.36 | 0.32 | 0.48 | 0.33 | 0.24 | 0.32 | 0.35 (0.08) |
| 11 | 0.28 | 0.17 | 0.56 | 0.53 | 0.23 | 0.49 | 2.60 | 0.13 | 0.10 | 0.57 (0.74) |
| 12 | 0.27 | 0.14 | 0.22 | 0.40 | 0.32 | 0.26 | 0.36 | 0.40 | 0.45 | 0.31 (0.09) |
| 13 | 0.77 | 0.71 | 0.58 | 0.77 | 0.53 | 0.61 | 0.87 | 0.67 | 0.55 | 0.67 (0.11) |
| 14 | 0.36 | 0.44 | 0.28 | 0.48 | 0.17 | 0.42 | 0.27 | 0.13 | 0.31 | 0.32 (0.11) |
| 15 | 1.24 | 0.62 | 1.88 | 1.07 | 0.60 | 0.62 | 0.57 | 0.92 | 1.12 | 0.96 (0.40) |
| 16 | 0.40 | 0.58 | 0.50 | 0.36 | 0.40 | 0.81 | 0.20 | 0.41 | 1.21 | 0.54 (0.28) |
| 17 | 0.73 | 0.48 | 0.39 | 0.69 | 0.49 | 0.18 | 4.63 | 0.46 | 0.51 | 0.95 (1.31) |
| 18 | 0.32 | 0.22 | 0.16 | 0.18 | 0.28 | 0.34 | 0.42 | 0.26 | 3.77 | 0.66 (1.10) |
| 19 | 0.54 | 0.57 | 0.65 | 0.53 | 0.53 | 0.68 | 0.47 | 0.61 | 0.58 | 0.57 (0.06) |
| 20 | 0.17 | 0.50 | 0.25 | 0.24 | 0.17 | 0.22 | 0.26 | 0.37 | 0.27 | 0.27 (0.10) |
| 21 | 0.10 | 0.24 | 0.23 | 0.13 | 0.11 | 0.24 | 0.29 | 0.28 | 0.30 | 0.21 (0.07) |
| 22 | 1.29 | 0.68 | 0.86 | 1.21 | 0.93 | 0.85 | 1.12 | 1.28 | 0.84 | 1.01 (0.21) |
| 23 | 0.67 | 0.24 | 1.16 | 0.81 | 0.28 | 0.33 | 0.38 | 8.55 | 2.28 | 1.63 (2.52) |
| 24 | 0.47 | 0.57 | 0.51 | 0.40 | 0.37 | 0.61 | 0.53 | 0.42 | 0.41 | 0.48 (0.08) |
| 25 | 0.50 | 4.31 | 0.96 | 3.03 | 0.40 | 0.47 | 0.41 | 0.32 | 0.58 | 1.22 (1.35) |
| 26 | 0.28 | 0.34 | 0.51 | 0.33 | 0.30 | 0.51 | 0.36 | 0.40 | 0.52 | 0.39 (0.09) |
| Mean (SD) | 0.56 (0.43) | 0.86 (1.25) | 0.65 (0.49) | 0.62 (0.57) | 0.43 (0.22) | 0.55 (0.36) | 0.78 (0.96) | 0.89 (1.65) | 1.00 (1.6) | |

*2nd round*

*Unity pipeline*

| | | | | | | | | | | |
|---|---|---|---|---|---|---|---|---|---|---|
| 01 | 0.33 | 0.25 | 0.29 | 0.22 | 0.16 | 0.23 | 0.17 | 0.22 | 0.33 | 0.25 (0.06) |
| 02 | 0.46 | 0.48 | 0.89 | 0.77 | 0.81 | 0.81 | 0.52 | 0.97 | 0.80 | 0.72 (0.18) |
| 03 | 0.49 | 0.68 | 0.52 | 0.35 | 0.56 | 0.46 | 0.40 | 0.49 | 0.51 | 0.50 (0.09) |
| 04 | 0.70 | 0.99 | 1.88 | 1.03 | 0.53 | 1.16 | 0.57 | 0.45 | 0.84 | 0.90 (0.41) |
| 05 | 0.88 | 5.31 | 2.40 | 0.41 | 2.20 | 0.88 | 0.46 | 0.37 | 0.52 | 1.49 (1.53) |
| 06 | 0.27 | 0.25 | 0.25 | 0.45 | 0.23 | 0.15 | 0.29 | 0.20 | 0.73 | 0.31 (0.17) |
| 07 | 0.63 | 1.38 | 1.49 | 0.90 | 1.11 | 1.40 | 0.99 | 0.93 | 1.29 | 1.13 (0.27) |
| 08 | 10.07 | 5.00 | 5.73 | 8.57 | 2.55 | 2.71 | 6.51 | 1.35 | 9.20 | 5.74 (2.95) |
| 09 | 0.41 | 0.36 | 0.67 | 0.33 | 0.41 | 0.23 | 0.18 | 0.20 | 0.16 | 0.33 (0.15) |
| 10 | 4.28 | 0.16 | 9.46 | 0.28 | 0.25 | 0.32 | 0.28 | 0.25 | 0.40 | 1.74 (3.00) |
| 11 | 0.49 | 0.50 | 0.49 | 7.64 | 0.36 | 0.39 | 0.57 | 0.55 | 0.45 | 1.27 (2.25) |
| 12 | 0.83 | 0.71 | 0.67 | 0.74 | 0.67 | 1.36 | 0.62 | 0.25 | 0.50 | 0.71 (0.28) |
| 13 | 4.66 | 0.49 | 0.60 | 0.50 | 0.46 | 0.47 | 1.93 | 0.67 | 1.91 | 1.30 (1.32) |
| 14 | 0.64 | 0.74 | 1.33 | 0.35 | 0.19 | 0.43 | 0.30 | 0.17 | 0.49 | 0.51 (0.34) |
| 15 | 0.53 | 0.61 | 0.90 | 0.54 | 0.33 | 0.33 | 0.43 | 0.33 | 0.22 | 0.47 (0.19) |
| 16 | 1.66 | 0.57 | 4.85 | 1.69 | 5.61 | 4.99 | 6.92 | 2.18 | 10.61 | 4.34 (3.01) |
| 17 | 0.28 | 0.32 | 0.44 | 0.37 | 0.26 | 0.42 | 0.77 | 0.58 | 1.24 | 0.52 (0.30) |
| 18 | 0.88 | 0.99 | 0.53 | 4.00 | 0.84 | 0.68 | 0.83 | 0.67 | 8.54 | 1.99 (2.53) |
| 19 | 0.61 | 0.57 | 0.52 | 0.60 | 0.91 | 0.81 | 2.51 | 0.82 | 0.58 | 0.88 (0.59) |
| 20 | 0.39 | 0.23 | 0.32 | 1.97 | 0.18 | 0.23 | 0.33 | 0.17 | 0.67 | 0.50 (0.54) |
| 21 | 0.28 | 0.21 | 0.16 | 0.24 | 0.09 | 0.14 | 0.29 | 0.17 | 0.26 | 0.21 (0.06) |
| 22 | 0.69 | 0.76 | 1.64 | 0.62 | 0.72 | 1.52 | 0.97 | 0.57 | 1.19 | 0.96 (0.37) |
| 23 | 1.50 | 1.03 | 1.30 | 0.89 | 0.47 | 0.72 | 0.68 | 0.76 | 0.45 | 0.87 (0.34) |
| 24 | 0.50 | 0.53 | 0.49 | 0.33 | 0.40 | 0.33 | 0.59 | 0.39 | 0.28 | 0.43 (0.10) |
| 25 | 1.14 | 1.04 | 3.96 | 0.71 | 0.86 | 1.86 | 0.70 | 0.73 | 1.89 | 1.43 (0.99) |
| 26 | 0.32 | 0.38 | 0.24 | 0.45 | 0.33 | 0.44 | 0.30 | 0.46 | 0.49 | 0.38 (0.08) |
| Mean (SD | 1.31 (2.06) | 0.94 (1.25) | 1.62 (2.11) | 1.34 (2.10) | 0.83 (1.11) | 0.90 (1.01) | 1.12 (1.69) | 0.57 (0.43) | 1.71 (2.84) | |

Table A3 The user ratings on the System Usability Scale

| No. | 1. I think that I would like to use AVP frequently. | 2. I found the AVP unnecessarily complex. | 3. I thought the AVP was easy to use. | 4. I think that I would need the support of a technical person to be able to use the AVP. | 5. I found the various functions in the AVP were well integrated. | 6. I thought there was too much inconsistency in the AVP | 7. I would imagine that most people would learn to use AVP very quickly. | 8. I found AVP very awkward to use. | 9. I felt very confident using the AVP | 10. I needed to learn a log of thins before I could get going with AVP. | Usable | Learnable | Overall |
|---|---|---|---|---|---|---|---|---|---|---|---|---|---|
| | **SUS items** | | | | | | | | | | **Standardized SUS score** | | |
| 01 | 5 | 2 | 3 | 5 | 5 | 3 | 3 | 2 | 5 | 4 | 75.00 | 12.50 | 62.50 |
| 02 | 3 | 1 | 5 | 1 | 3 | 2 | 5 | 2 | 5 | 1 | 81.25 | 100.00 | 85.00 |
| 03 | 5 | 2 | 4 | 2 | 4 | 2 | 4 | 2 | 4 | 3 | 78.13 | 62.50 | 75.00 |
| 04 | 5 | 3 | 4 | 3 | 4 | 4 | 4 | 3 | 4 | 2 | 65.63 | 62.5 | 65.00 |
| 05 | 4 | 2 | 4 | 2 | 4 | 2 | 4 | 2 | 4 | 2 | 75.00 | 75.00 | 75.00 |
| 06 | 4 | 3 | 5 | 4 | 3 | 2 | 4 | 2 | 4 | 2 | 71.88 | 50.00 | 67.50 |
| 07 | 5 | 2 | 3 | 3 | 4 | 1 | 3 | 2 | 5 | 2 | 78.13 | 62.50 | 75.00 |
| 08 | 5 | 2 | 4 | 3 | 4 | 3 | 4 | 2 | 5 | 2 | 78.13 | 62.50 | 75.00 |
| 09 | 4 | 2 | 3 | 3 | 3 | 1 | 5 | 1 | 5 | 1 | 81.25 | 75.00 | 80.00 |
| 10 | 3 | 2 | 4 | 2 | 5 | 3 | 5 | 2 | 5 | 3 | 78.13 | 62.50 | 75.00 |
| 11 | 4 | 2 | 4 | 2 | 3 | 2 | 4 | 2 | 5 | 2 | 75.00 | 75.00 | 75.00 |
| 12 | 5 | 2 | 4 | 1 | 5 | 4 | 5 | 1 | 5 | 1 | 84.38 | 100.00 | 87.50 |
| 13 | 3 | 2 | 4 | 3 | 3 | 3 | 4 | 2 | 4 | 2 | 65.63 | 62.50 | 65.00 |
| 14 | 5 | 3 | 3 | 3 | 4 | 4 | 4 | 2 | 4 | 4 | 65.63 | 37.50 | 60.00 |
| 15 | 4 | 2 | 4 | 2 | 4 | 2 | 4 | 2 | 4 | 3 | 75.00 | 62.50 | 72.50 |
| 16 | 5 | 1 | 3 | 3 | 4 | 2 | 3 | 3 | 3 | 2 | 68.75 | 62.50 | 67.50 |
| 17 | 3 | 1 | 5 | 3 | 4 | 1 | 5 | 2 | 4 | 4 | 84.38 | 37.50 | 75.00 |
| 18 | 4 | 3 | 4 | 1 | 4 | 5 | 4 | 2 | 5 | 1 | 65.63 | 100.00 | 72.50 |
| 19 | 4 | 2 | 4 | 1 | 4 | 2 | 3 | 1 | 5 | 3 | 78.13 | 75.00 | 77.50 |
| 20 | 3 | 1 | 3 | 1 | 2 | 2 | 4 | 2 | 5 | 1 | 68.75 | 100.00 | 75.00 |
| 21 | 4 | 2 | 4 | 3 | 3 | 2 | 3 | 2 | 4 | 2 | 68.75 | 62.50 | 67.50 |
| 22 | 4 | 2 | 4 | 2 | 3 | 4 | 4 | 3 | 3 | 2 | 59.38 | 75.00 | 62.50 |
| 23 | 5 | 1 | 5 | 2 | 4 | 2 | 5 | 2 | 5 | 1 | 90.63 | 87.50 | 90.00 |
| 24 | 5 | 2 | 4 | 3 | 3 | 3 | 5 | 2 | 5 | 1 | 78.13 | 75.00 | 77.50 |
| 25 | 3 | 1 | 5 | 1 | 5 | 1 | 5 | 3 | 5 | 2 | 87.50 | 87.50 | 87.50 |
| 26 | 5 | 2 | 4 | 3 | 4 | 2 | 4 | 2 | 4 | 3 | 78.13 | 50.00 | 72.50 |